\documentclass[5p]{elsarticle}

\usepackage{hyperref}
\usepackage{lineno}
\modulolinenumbers[5]

\usepackage{graphicx,color}
\usepackage{amsmath}
\usepackage{epstopdf}
\usepackage{verbatim}
\usepackage{color}
\usepackage{soul}
\setstcolor{red}

\usepackage[e]{esvect}

\newcommand{\al}[1]{\textcolor{black}{#1}}

\newcommand{\all}[1]{\textcolor{black}{#1}}

\journal{Journal of Magnetic Resonance}

\begin{document}

\begin{frontmatter}

\title{Cantilever detected ferromagnetic resonance in thin Fe$_{50}$Ni$_{50}$, Co$_2$FeAl$_{0.5}$Si$_{0.5}$ and Sr$_2$FeMoO$_6$ films using a double modulation technique.}


\author[1]{Alexey Alfonsov\corref{mycorrespondingauthor}}
\cortext[mycorrespondingauthor]{Corresponding author}
\ead{a.alfonsov@ifw-dresden.de}

\author[2]{Eiji Ohmichi}
\author[3]{Pavel Leksin}
\author[3]{Ahmad Omar}
\author[4]{Hailong Wang}
\author[3,5]{Sabine Wurmehl}
\author[4]{Fengyuan Yang}
\author[1,2]{Hitoshi Ohta}

\address[1]{Molecular Photoscience Research Center, Kobe University, Kobe 657-8501, Japan}
\address[2]{Graduate School of Science, Kobe University, 1-1 Rokkodai-cho,
Nada, Kobe 657-8501, Japan}
\address[3]{Leibniz Institute for Solid State and Materials Research Dresden, IFW, D-01171 Dresden, Germany}
\address[4]{Department of Physics, The Ohio State University, Columbus, Ohio 43210, USA}
\address[5]{Institut fur Festk\"orperphysik, Technische Universit\"at Dresden, D-01062 Dresden, Germany}

\begin{abstract}
In this work we introduce a new method of a ferromagnetic resonance (FMR) detection from thin, nm-size, films. Our setup is based on the commercial piezo-cantilever, used for atomic force microscopy. It has an option to rotate the sample in the magnetic field and it operates up to the high microwave frequencies of 160 GHz. Using our cantilever based FMR spectrometer we have investigated a set of samples, namely quasi-bulk and 84 nm film Co$_2$FeAl$_{0.5}$Si$_{0.5}$ samples, 16 nm Fe$_{50}$Ni$_{50}$ film and 150 nm Sr$_2$FeMoO$_6$ film. The high frequency ferromagnetic resonance (FMR) response from an extremely thin Fe$_{50}$Ni$_{50}$ film we have fitted with the conventional model for the magnetization dynamics. The cantilever detected FMR experiments on Sr$_2$FeMoO$_6$ film reveal an inability of the conventional model to fit frequency and angular dependences with the same set of parameters, which suggests that one has to take into account much more complicated nature of the magnetization precession in the  Sr$_2$FeMoO$_6$ at low temperatures and high frequencies. Moreover, the complicated dynamics of the magnetization apparent in all investigated samples is suggested by a drastic increase of the linewidths with increasing microwave frequency, and by an emergence of the second line with an opposite angular dependence.
\end{abstract}


\end{frontmatter}

%\linenumbers

\section{Introduction}

A key property of many intensively studied materials is a magnetic anisotropy, which is defined by the complex interplay of  different degrees of freedom, such as spin or/and orbital moments, charge and lattice. In particular, in the  ferromagnets promising for spintronic application, such as Sr$_2$FeMoO$_6$ \cite{KKS98, RAM04} and Co$_2$FeAl$_{0.5}$Si$_{0.5}$ \cite{TIS07, FFG07}, magnetic anisotropy defines the thermal stability of the magnetization. One of the most appropriate methods to study magnetic anisotropies, as well as gyromagnetic ratios and magnetization dynamics in ferromagnets is ferromagnetic resonance (FMR). 

Potential halfmetallic ferromagnetic materials has been already studied by means of ferromagnetic resonance. In most cases the study was limited to room temperatures and low frequencies of standard spectrometers \cite{YOA07, NML2008,Du2013}, whereas there are also measurements performed using vector network analysers and microstrip resonators at frequencies up to 70 GHz \cite{ZLB07, MWO09, GBP15}. Additionally there are few reports of measurements at even higher frequencies \cite{OIM93}. Increasing a measurement frequency in the FMR experiment is very important, since it yields a higher resolution and, therefore, better determination of g-factors and magnetic anisotropies. Unfortunately high-frequency measurements are associated with several problems in the detection of the FMR signal, especially when performing measurements on thin, nm-size, films. Namely, in order to increase the sensitivity one has to apply restrictions on the microwave frequency, or strength and orientation of magnetic field. For instance, standard resonators that amplify the microwave power at the sample and drastically increase the sensitivity, are often used only in a narrow frequency range.

In this work first we introduce a new method of detecting ferromagnetic resonance from thin, nm-size, films, where all the restrictions named above are lifted. Our setup, which is based on the measurement of the deflection of a $\mu$m-size piezo-cantilever, has an option to change the angle between magnetic field and the film plane, and it works up to the high microwave frequencies of 160 GHz. Using this setup we investigated a set of selected \al{materials}, namely quasi-bulk Co$_2$FeAl$_{0.5}$Si$_{0.5}$, 84 nm Co$_2$FeAl$_{0.5}$Si$_{0.5}$ film, 16 nm Fe$_{50}$Ni$_{50}$ film and 150 nm Sr$_2$FeMoO$_6$ film. We have measured high frequency ferromagnetic resonance response from an extremely thin Fe$_{50}$Ni$_{50}$ film and fitted it with the conventional model for the magnetization dynamics \cite{Smit1955, Farle1998}. The cantilever detected FMR experiments on Sr$_2$FeMoO$_6$ film reveal an inability of the conventional model to fit frequency and angular dependences with the same set of parameters, which suggests that one has to take into account much more complicated nature of the magnetization precession in the  Sr$_2$FeMoO$_6$ at low temperatures and high frequencies. Moreover, the complicated dynamics of the magnetization apparent in all the investigated samples is supported by a drastic increase of the linewidths with increasing microwave frequency, and by the emergence of the second line with an opposite angular dependence.

\section{Investigated samples}

\al{Polycrystalline Co$_2$FeAl$_{0.5}$Si$_{0.5}$ bulk sample was cast by arc-melting stoichiometric quantities of at least 4N pure constituents. Before melting the sample itself, the chamber was evacuated to $10^{-5}$ mbar pressure before backfilling with argon followed by melting a Ti piece in order to minimize oxygen. In order to achieve good homogeneity, the sample was flipped and re-melted 4 times. Resulting cast sample was then sealed in an evacuated quartz ampoule and subsequently annealed at 1400 K for 3 days followed by slow-cooling. The annealed sample was mechanically grinded to a thin plate of $\sim$17 $\mu$m.}

Epitaxial 84 nm Co$_2$FeAl$_{0.5}$Si$_{0.5}$ film was grown on
MgAl$_2$O$_4$ (001) substrate by off-axis sputtering in a UHV
system with a base pressure as low as 9.5$\cdot$10$^{-11}$ mbar
using ultra-pure Ar (99.9999\%) as sputtering gas. Optimal quality
Co$_2$FeAl$_{0.5}$Si$_{0.5}$ epitaxial film was obtained at an
Ar pressure of 6$\cdot$10$^{-3}$ mbar, a substrate temperature of 600$^\circ$C,
and DC sputtering at a constant current of 12~mA, which results in
a deposition rate of 5.6 \AA/min \cite{PAB13,APY15}.

The deposition of 16 nm Fe$_{50}$Ni$_{50}$ film was performed using an e-gun in ultra high vacuum (UHV) with pressure $10^{-9}$ mbar on the MgO(100) substrate  at a temperature T$_{\mathrm{sub}}$ = 300 K \cite{Leksin2015}. The deposition rate was set to be 30 \AA/min.

Epitaxial Sr$_2$FeMoO$_6$ films of 150 nm thickness were grown on SrTiO$_3$ (001) substrates in an ultrahigh vacuum sputtering system with a base pressure of $6.5\cdot10^{-10}$ mbar using a stoichiometric Sr$_2$FeMoO$_6$ targets \cite{Hauser2011, Du2013}. Direct-current (DC) magnetron sputtering was used for film deposition with a constant current of 5 mA in a pure Ar gas (99.9995\%) of 9$\cdot$10$^{-3}$ mbar, which resulted in a growth rate of 0.84 \AA/min.  The films were deposited in 90\,$^o$ off-axis geometry with the substrate temperature maintained at 800$^\circ$C.

\section{Experimental setup}

%%%%%%%%%%%%%%%%%%%%%%%%%%%%%%%%%%%%%%%%
\begin{figure}
\centering
\includegraphics[width=8cm]{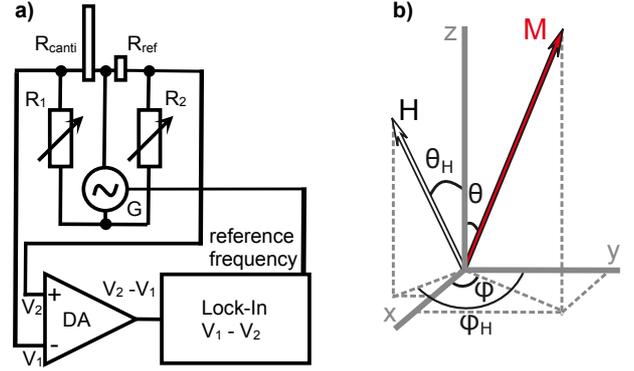}
\caption{a) Schematic diagram of a piezocantilever resistance detection system based on the Wheatstone bridge. b) Definition of the magnetization and the magnetic field directions with x-y being a film plane. }
\label{Fig_setup}
\end{figure}
%%%%%%%%%%%%%%%%%%%%%%%%%%%%%%%%%%%%%%%%

\al{The present experimental setup is based on the cantilever detected ESR setup developed in the Molecular Photoscience Research Center in Kobe University, Kobe, Japan \cite{Ohmichi8, Ohmichi2010, Ohmichi2013, TOO15}}. The sample is glued directly to a commercial piezo-cantilever \cite{Ohmichi2010, TOO15,  OO2002}. The deflection of the cantilever is measured by means of a bias box based on the Wheatstone bridge, alternating voltage generator (G) and differential amplifier (DA) (Fig.~\ref{Fig_setup}(a)). The alternating voltage generator (G) provides a voltage which feeds the bridge. R$_{canti}$ is the cantilever with the sample, whereas R$_{ref}$ is a reference unloaded cantilever \cite{Ohmichi8, Ohmichi2010}, which is needed to cancel unwanted temperature and magnetic field effects. Two variable resistances R$_1$ and R$_2$ are tuned in order to equalize all the bridge resistances so that the voltage difference V$_2$-V$_1$ is equal to zero. When the ESR or FMR absorption occurs, the cantilever \all{is pulled by the sample}, which leads to a change of R$_{canti}$, and therefore to a change of the difference V$_2$-V$_1$. This change is detected by a highly sensitive Lock-In amplifier, which tracks the signal at the reference frequency equal to a frequency of the voltage generator G (Fig.~\ref{Fig_setup}(a)). Our experiments showed that this frequency for the voltage modulation has to be near the half of the eigenfrequency of the cantilever to maximize the signal to noise ratio.

\begin{figure}
\centering
\includegraphics[width=8cm]{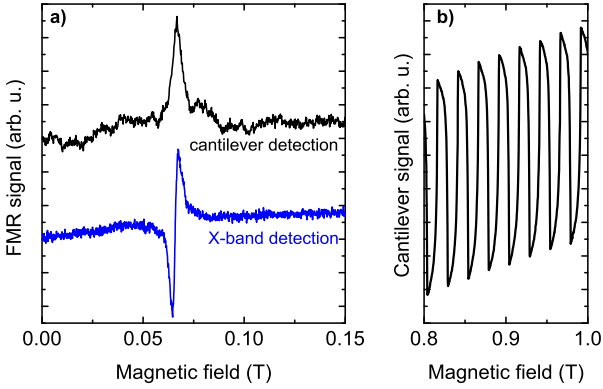}
\caption{\textbf{a)} Comparison of the FMR signal from 84 nm film of Co$_2$FeAl$_{0.5}$Si$_{0.5}$ measured at room temperature by the commercial X-Band spectrometer from Bruker (bottom line) and by our custom made setup with a piezocantilever acting as a detector (top line). The measurements were carried out simultaneously. \textbf{b)} Signal from the cantilever as the function of the magnetic field measured by double modulation technique. Modulated parameters are the frequency of the voltage applied to the Wheatstone bridge and the power of the microwave.}
\label{Fig_xband}
\end{figure}

In order to check the sensitivity of the cantilever based spectrometer we performed the following experiment. The 84 nm Co$_2$FeAl$_{0.5}$Si$_{0.5}$ thin film sample glued to a cantilever was inserted into a cavity of the highly sensitive commercial Bruker X-Band (9.56 GHz) EPR spectrometer. In this experiment, the signal at the X-Band spectrometer detector was recorded together with the signal from piezocantilever while sweeping the magnetic field at room temperature. As can be seen in Fig.~\ref{Fig_xband}(a) the experiment showed that both detection systems yield identical FMR responses, and that the sensitivity of the piezocantilever detection system is comparable to that of the X-Band spectrometer from Bruker. Note that the Bruker X-Band spectrometer is measuring the absorption derivative, whereas the cantilever deflection represents a pure absorption, which is depicted in Fig.~\ref{Fig_xband}(a).

%%%%%%%%%%%%%%%%%%%%%%%%%%%%%%%%%%%%%%%%

%%%%%%%%%%%%%%%%%%%%%%%%%%%%%%%%%%%%%%%%

It is important to bear in mind that during the cantilever detected FMR experiment we also measure a deflection of the cantilever due to the magnetic anisotropy of the sample. This is a very strong non-resonance effect, which depends on the sample magnetization and the angle between film plane and the applied magnetic field. Such cantilever deflection is the same with or without the microwave radiation. Therefore there is a need to subtract the background signal from such non-resonant magnetic response. First option is to measure the cantilever response twice, first time with the microwave source switched on and second time with it switched off. Then the latter measurement can be subtracted from the former one to get the pure FMR signal. In this case the drawback is that any drift of the background during both measurements will be seen in the result of the subtraction. To avoid this and to make these two measurements quasi-simultaneous, we have implemented a second modulation, in this case the modulation of the microwave power (on/off) at a low frequency of $\sim0.2$ Hz. The raw measurement data are shown in Figure~\ref{Fig_xband}(b). As can be seen the signal is oscillating with the second modulation frequency ($\sim0.2$ Hz), and the pure FMR response can be obtained by subtracting the lower envelope of this signal from the upper one.

\begin{comment}
%%%%%%%%%%%%%%%%%%%%%%%%%%%%%%%%%%%%%%%%
\begin{figure}
\centering
\includegraphics[width=8cm]{sensitivity.eps}
\caption{\al{not sure if we should show it}}
\label{Fig_sensitivity}
\end{figure}
%%%%%%%%%%%%%%%%%%%%%%%%%%%%%%%%%%%%%%%%
\end{comment}

High microwave frequency radiation was generated by a set of Gunn oscillators from Millitech. The cantilever was mounted on the probehead with \textit{in-situ} sample rotation mechanism, described elsewhere \cite{Ohmichi2013}. The probehead was inserted into a magneto-cryostat from Oxford Instruments with maximum magnetic field of 15 T. The temperature of all the high frequency experiments was 8 K.

\section{Model for analysis of the FMR results}
\label{sec_model}

To analyze the frequency and the angular dependences of the measured spectra we have used a standard approach \cite{Smit1955, Farle1998}, where the resonance frequency is given by the following equation:

\begin{align}
\omega^2 = \frac{\gamma^2}{M_s^2\sin^2\theta} \bigg(\frac{\partial^2\mathbf{E}}{\partial \theta ^2}\frac{\partial^2\mathbf{E}}{\partial \phi^2} - \Big(\frac{\partial^2\mathbf{E}}{\partial \theta \partial \phi}\Big)^2\bigg)
\label{eq_f}
\end{align}

Here $\gamma$ is the gyromagnetic ratio and the energy density \textbf{E} is shown below:

\begin{align}
\mathbf{E} = - (\vv{\textbf{H}} \cdot \vv{\textbf{M}}) + 2\pi M_s^2\cos^2\theta - K_U\cos^2\theta \notag \\ + K_c(\alpha_x^2\alpha_y^2 + \alpha_x^2\alpha_z^2 + \alpha_y^2\alpha_z^2)
\label{eq_E}
\end{align}
where
\begin{align}
\alpha_x = \sin\theta\cos\phi \notag; \ \ \
\alpha_y = \sin\theta\sin\phi \notag; \ \ \
\alpha_z = \cos\theta \notag
\end{align}

\begin{comment}
\begin{align}
\alpha_x &= \sin\theta\cos\phi \notag \\
\alpha_y &= \sin\theta\sin\phi \notag \\
\alpha_z &= \cos\theta \notag
\end{align}
\end{comment}

In Eq.\ref{eq_E} the first term represents the Zeeman energy with $\vv{\mathbf{H}}(H, \theta_H, \phi_H)$ being the applied magnetic field and $\vv{\mathbf{M}}(M_s, \theta, \phi)$ being the magnetization, M$_s$ is the saturation magnetization. Second term is a shape anisotropy energy for a thin film. Third and fourth terms are uniaxial and cubic anisotropies with $K_U$ and $K_c$ being the \al{out-of-plane} uniaxial anisotropy and cubic anisotropy constants, respectively. $\theta$ ($\theta_H$) is the angle between the perpendicular to the film plane and the magnetization direction (applied magnetic field $\vv{\mathbf{H}}$). $\phi$ ($\phi_H$) is the angle which defines the direction of the magnetization (applied magnetic field $\vv{\mathbf{H}}$) in the film plane (see Fig.~\ref{Fig_setup}(b)). External magnetic field $\vv{\textbf{H}}$ is defined as it is applied in the experiment, and the angles $\theta$ and $\phi$ are found by numerical minimization of the energy density (Eq. \ref{eq_E}) for each given condition. As the result of the fitting the Eq. \ref{eq_f} to the experimental data we obtain $\gamma$, $K_U$ and $K_c$.

\begin{comment}
%%%%%%%%%%%%%%%%%%%%%%%%%%%%%%%%%%%%%%%%
\begin{figure}
\centering
\includegraphics[width=2cm]{angles.eps}
\caption{angles}
\label{Fig_angles}
\end{figure}
%%%%%%%%%%%%%%%%%%%%%%%%%%%%%%%%%%%%%%%%
\end{comment}

\section{Experimental results}
\label{sec_res}

\subsection{Overview of the high-frequency FMR response from all investigated samples}

%%%%%%%%%%%%%%%%%%%%%%%%%%%%%%%%%%%%%%%%
\begin{figure}
\centering
\includegraphics[width=8cm]{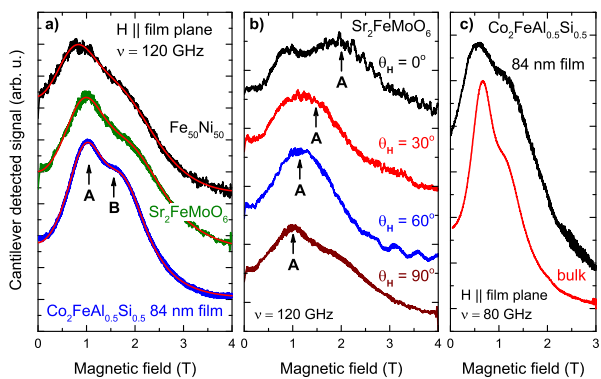}
\caption{a) Comparison of the cantilever detected FMR spectra of Co$_2$FeAl$_{0.5}$Si$_{0.5}$ film (bottom line), Fe$_{50}$Ni$_{50}$ film (top line) and Sr$_2$FeMoO$_6$ film (middle line), measured at the temperature of 8 K, at the frequency of 120 GHz and with magnetic field lying in the films plane. The solid lines over measured spectra represent the two Lorentzian lines fit. b) Angular dependence of the cantilever detected FMR signal measured at the temperature of 8 K and at the frequency of 120 GHz. $\theta_H$ is the angle between perpendicular of the film and the applied magnetic field. c) Comparison  of the cantilever detected FMR spectra of Co$_2$FeAl$_{0.5}$Si$_{0.5}$ bulk (17 $\mu$m thick plate) sample (bottom line) and Co$_2$FeAl$_{0.5}$Si$_{0.5}$ film (top line) measured at the temperature of 8 K and at the frequency of 80 GHz.}
\label{Fig_spectra_comp}
\end{figure}
%%%%%%%%%%%%%%%%%%%%%%%%%%%%%%%%%%%%%%%%

As can be seen in Fig. \ref{Fig_spectra_comp}(a) all the investigated thin films exhibit a rather strong FMR response at 8 K and at the measurement microwave frequency of 120 GHz. All the FMR signals consist of 2 lines, low (\textbf{A}) and high (\textbf{B}) field peaks. At this frequency lines are rather broad, with the widths ranging from $\sim$1 T to $\sim$1.5 T for the first peak, and from $\sim$1.2 T to $\sim$2.1 T for the second peak, respectively. These values were obtained by fitting FMR signals with two Lorentzian lines, respective fits are shown in the same Fig. \ref{Fig_spectra_comp}(a) as solid lines. \al{Interestingly, the linewidths measured at high frequencies using cantilever detection are much larger then those measured at low frequencies using Bruker X-Band spectrometer, this is especially noticeable in the case of Fe$_{50}$Ni$_{50}$ film (see spectra in Fig. \ref{Fig_feni}). Such a drastic increase of the linewidths with increasing microwave frequency at low temperatures remains an open question, but most likely points to the complicated dynamics of the magnetization apparent in all the investigated samples. As has been shown, the linewidth in the case of FMR is defined by the complex interplay of different mechanisms, including Gilbert damping, mosaicity, two magnon scaterring, etc. (Ref. \cite{ZLB07, Hei2005} and references therein), and therefore needs a further investigations without restrictions in frequencies and temperatures.}

The angular dependence of the FMR response shows a peculiar behavior for the second line \textbf{B} (Fig. \ref{Fig_spectra_comp}(b)). Whereas the line \textbf{A} shifts to higher fields, as expected for FMR signal when magnetic field is rotated from the plane of the film ($\theta_H = 90~^\mathrm{o}$, Fig.~\ref{Fig_setup}(b)) towards perpendicular direction ($\theta_H = 0~^\mathrm{o}$), the second one shifts to lower fields. Therefore the more intensive first line (\textbf{A})\al{, or at least its resonance position,} we can attribute to a uniform precession mode, and the second line (\textbf{B}) is another mode excited in the sample, possibly perpendicular standing spin wave mode \cite{Belm2009} excited at such high frequencies. Due to the unclear nature of the high field peak \textbf{B}, we will focus on the analysis of the low field line \textbf{A} with expected angular dependence.

Comparing the Co$_2$FeAl$_{0.5}$Si$_{0.5}$ thin film FMR signal\footnotemark[1] to the one from a quasi-bulk ($17~\mu$m - thick) sample (see Fig. \ref{Fig_spectra_comp}(c)), we see that qualitatively the shape of the FMR signal  remains the same, although the details, such as resonance positions and the linewidths of the individual lines \textbf{A} and \textbf{B} are different. Interestingly, the linewidth is somewhat smaller in the case of $17~\mu$m - thick plate. This suggests that such double-line shape is not only due to a small thickness of the measured films, but also represents an intrinsic effect related to the high frequency of the excitation. As can be seen in Fig. \ref{Fig_sfmo}(b) in the case of Sr$_2$FeMoO$_6$ sample, the frequency when the second line becomes apparent should be above $\sim60$ GHz.

\footnotetext[1]{The study of the Co$_2$FeAl$_{0.5}$Si$_{0.5}$ thin film samples with different thickness is published elsewhere.}

\subsection{FMR measurements on Fe$_{50}$Ni$_{50}$ 16 nm film}

%%%%%%%%%%%%%%%%%%%%%%%%%%%%%%%%%%%%%%%%
\begin{figure}
\centering
\includegraphics[width=8cm]{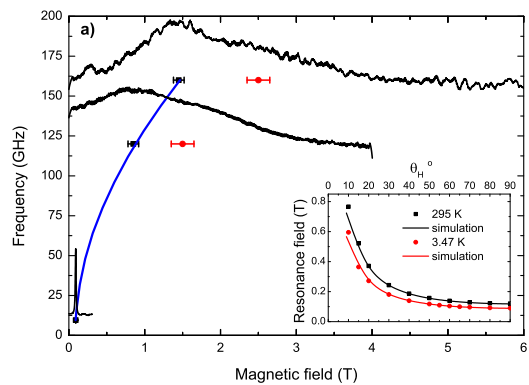}
\caption{Frequency as a function of the resonance field for Fe$_{50}$Ni$_{50}$ 16 nm film, shown together with measured spectra. Spectra measured at 120 GHz and 160 GHz were detected by the cantilever deflection. Spectrum at 9.56 GHz was measured using Bruker X-band spectrometer, here the integral is shown. The angular dependences of the resonance fields measured at 9.56 GHz together with the simulated curves are shown in the inset.}
\label{Fig_feni}
\end{figure}
%%%%%%%%%%%%%%%%%%%%%%%%%%%%%%%%%%%%%%%%

In the case of Fe$_{50}$Ni$_{50}$ film two spectra were measured at 120 GHz and 160 GHz at the temperature of 8 K and with external magnetic field lying in the film plane (Fig. \ref{Fig_feni}). In addition two angular dependences were measured by Bruker X-Band spectrometer at 9.56 GHz, one at room temperature, and another at 3.47 K (see inset in Fig. \ref{Fig_feni}). Fitting the high frequency FMR response with two Lorentzian lines we were able to extract the resonance positions of both peaks, which are shown together with respective spectra in Fig. \ref{Fig_feni}. Using the model described in the Section \ref{sec_model}, we have fitted the frequency vs resonance field dependence for the magnetic field being in the film plane and the angular dependence measured at the X-band frequency of 9.56 GHz and T = 3.47~K simultaneously. The fit, which is shown in Fig. \ref{Fig_feni} as solid lines, yields $\gamma/2\pi = 28$ GHz/T, cubic anisotropy constant $K_c = - 5\cdot10^5$ erg\,cm$^{-3}$ and uniaxial anisotropy constant $K_U = - 125\cdot10^6$ erg\,cm$^{-3}$. The negative sign and significant value of $K_U$ points to the fact that Fe$_{50}$Ni$_{50}$ has an easy plane anisotropy, which can be explained by the noticeable lattice mismatch between Fe$_{50}$Ni$_{50}$ ($a = 3.578$ \cite{KZP2010}) and MgO substrate ($a = 4.212$ \cite{SL68}), especially considering a rather small film thickness of 16 nm. Importantly, reducing only the uniaxial anisotropy constant to the vale of  $K_U = - 10\cdot10^6$ erg\,cm$^{-3}$ in our simulations, we can fit the room temperature angular dependence measured at the frequency of 9.56 GHz. Such drastic increase of the anisotropy constants with decreasing temperature has been reported before \cite{Liu2003} \al{and explained by the collective change of the lattice parameters and the total magnetization with changing temperature}.

\subsection{FMR measurements on Sr$_2$FeMoO$_6$ 150 nm film}

%%%%%%%%%%%%%%%%%%%%%%%%%%%%%%%%%%%%%%%%
\begin{figure}
\centering
\includegraphics[width=8cm]{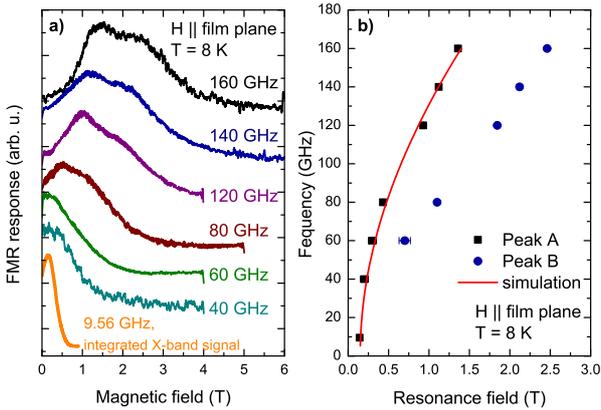}
\caption{a) Frequency dependence of the Sr$_2$FeMoO$_6$ FMR response, measured at $T = 8 K$ and with magnetic field lying in the film plane. The lowest line represents an integrated spectrum measured in the Bruker X-band spectrometer, the other lines are the spectra measured by the cantilever deflection. b) Frequency as a function of the resonance field for two Lorentzian lines constituting the spectrum (Squares for peak \textbf{A} and circles for peak \textbf{B}). Solid line represents a fit using model described in the section \ref{sec_model} (see text for details).}
\label{Fig_sfmo}
\end{figure}
%%%%%%%%%%%%%%%%%%%%%%%%%%%%%%%%%%%%%%%%

Besides the angular dependence measured at a frequency of 120 GHz and temperature of 8 K (see Fig. \ref{Fig_spectra_comp}(b)) we have measured a frequency dependence of the FMR signal with magnetic field lying in the film plane (Fig.\ref{Fig_sfmo}(a)). The fit with two Lorentzian lines described before yields a frequency vs resonance field diagram for both peaks, shown in Fig.\ref{Fig_sfmo}(b). Using the model described in the Section \ref{sec_model}, we have fitted this frequency vs resonance field dependence of the low field peak. The result of the fit is depicted as solid line. The saturation magnetization value for the equations \ref{eq_f} and \ref{eq_E} was taken from the static magnetometry measurements and is equal $M_s = 1.5 \mu$B/F.U. This fit yields the gyromagnetic ratio $\gamma/2\pi = 29.4$ GHz/T, cubic anisotropy constant $K_c = - 1.5\cdot10^5$ erg\,cm$^{-3}$ and uniaxial anisotropy constant $K_U = - 22.5\cdot10^6$ erg\,cm$^{-3}$. The $\gamma$ value is typical for Sr$_2$FeMoO$_6$ and agrees well with previous reports \cite{NML2008}. 

\begin{figure}
\centering
\includegraphics[width=8cm]{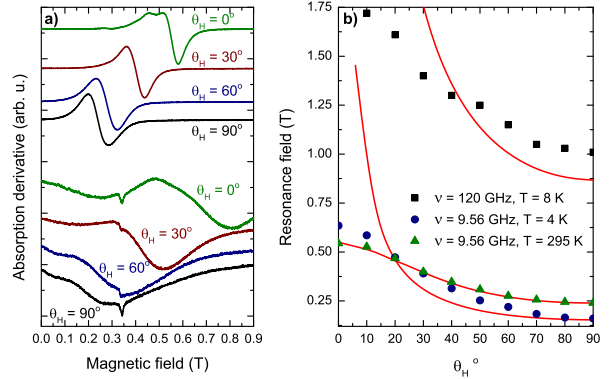}
\caption{a) Sr$_2$FeMoO$_6$ spectra measured in Bruker X-band spectrometer at $\nu = 9.56$ GHz, at different angles ($\theta_H$) between film plane and applied magnetic field. Four bottom lines represent the measurements performed at $T = 4$ K, four top lines are the ones at $T = 295$ K. b) The angular dependences of the resonance fields (symbols) with fits (solid lines) using model described in the section \ref{sec_model} (see text for details). Squares represent the measurement performed at $\nu = 120$ GHz, at $T = 8$ K using cantilever based setup. Circles and triangles represent the measurement using Bruker X-band spectrometer at $T = 4$ K and at $T = 295 K$, respectively.}
\label{Fig_sfmo_ang}
\end{figure}

In addition, we have measured the angular dependences of the FMR response at two temperatures, $T = 4$ K and $T = 295$ K, using commercial Bruker X-band spectrometer ($\nu = 9.56$ GHz). Representative spectra are shown in Fig. \ref{Fig_sfmo_ang}(a). The lines are rather broad, especially at $T = 4$ K. To obtain the resonance positions we have integrated all the spectra and picked all the resonance fields at the maxima of these integrals. These resonance positions are shown in Fig. \ref{Fig_sfmo_ang}(b) as a function of the angle $\theta_H$ together with the resonance positions measured at $\nu = 120$ GHz using cantilever detected FMR setup.

Taking the $\gamma$ value and the anisotropy constants obtained from the frequency vs resonance field dependence fit, we have simulated the low temperature angular dependences measured at $\nu = 9.56$ GHz and $\nu = 120$ GHz. As can be seen in Fig. \ref{Fig_sfmo_ang}(b), the simulated curves reproduce rather well the measured dependences at the angles near $\theta_H = 90^{\circ}$, \al{but strongly deviate at smaller angles}. Interestingly, the angular dependence measured at $\nu = 9.56$ GHz, at the room temperature, can be almost perfectly fitted with the gyromagnetic ratio $\gamma/2\pi = 29.4$ GHz/T and uniaxial anisotropy constant $K_U = - 4\cdot10^6$ erg\,cm$^{-3}$, which is noticeably smaller then its low temperature value. Such inability to find common parameters to fit all the data, like in the case of  Fe$_{50}$Ni$_{50}$ film, suggests that the model has to be reconsidered, taking into account much more complicated nature of the magnetization precession in the Sr$_2$FeMoO$_6$ at low temperatures and high microwave frequencies. 

%%%%%%%%%%%%%%%%%%%%%%%%%%%%%%%%%%%%%%%%
%%%%%%%%%%%%%%%%%%%%%%%%%%%%%%%%%%%%%%%%

\section{Summary}

In the present work we have introduced a new method for the detection of the ferromagnetic resonance of the samples with the thickness from $17$ $\mu$m down to $16$ nm. It is based on the deflection of the cantilever, which occurs at the ferromagnetic resonance. In order to increase the sensitivity we have implemented a double modulation, namely, we modulate the voltage at the Wheatstone bridge and the output of the microwave source at the same time. Using this setup we have investigated a set of samples: quasi-bulk and 84 nm film Co$_2$FeAl$_{0.5}$Si$_{0.5}$ sample, 16 nm Fe$_{50}$Ni$_{50}$ film and 150 nm Sr$_2$FeMoO$_6$ film. Low frequency test of our setup using a standard Bruker X-Band spectrometer showed identical FMR signals detected by the cantilever deflection and by the Bruker X-band detector simultaneously. \al{We have measured a high frequency ferromagnetic resonance (FMR) response from an quite thin Fe$_{50}$Ni$_{50}$ film and successfully fitted the resonance positions of the low field FMR lines together with the low frequency measurements using the conventional model for the magnetization dynamics.} The cantilever detected FMR experiments on Sr$_2$FeMoO$_6$ film revealed an inability of the conventional model to perfectly fit both frequency and angular dependences of the FMR signal with the same values of $\gamma$, $K_U$ and $K_c$, which suggests that one has to take into account much more complicated nature of the magnetization precession in the  Sr$_2$FeMoO$_6$ at low temperatures and high frequencies. Moreover, the complicated dynamics of the magnetization apparent in all the investigated samples is supported by a drastic increase of the linewidths with increasing microwave frequency, and by the emergence of the second line with an opposite angular dependence. These large linewidths and the emergence of the second lines remain an open question.

\section*{Acknowledgment}

This research has been supported by a DFG international research grant, project numbers AL 1771/1-1 and AL 1771/2-1, and by the Emmy Noether WU595/3-1 and the materials world network WU595/5-1. We would like to thank Dr. T. Sakurai and Dr. S. Okubo for support during cantilever detected FMR experiments. We also express our gratitude to Dr. V. Kataev for fruitful discussions. 

\section*{References}

\bibliography{mybibfile}

\end{document}